\documentclass{ckm}                 % twocolumn proceedings style
\usepackage{txfonts}            % if you have it, to get Times family
\usepackage{psfrag}
\newcommand{\be}{\begin{equation}}
\newcommand{\ee}{\end{equation}}
\newcommand{\bea}{\begin{eqnarray}}
\newcommand{\eea}{\end{eqnarray}}
\def\lsim{\raise0.3ex\hbox{$\;<$\kern-0.75em\raise-1.1ex\hbox{$\sim\;$}}}
\def\gsim{\raise0.3ex\hbox{$\;>$\kern-0.75em\raise-1.1ex\hbox{$\sim\;$}}}
\def\Frac#1#2{\frac{\displaystyle{#1}}{\displaystyle{#2}}}

\def\ep{\eta^{\prime}}
\def\susy{\mbox{\tiny SUSY}}
\def\sm{\mbox{\tiny SM}}
\confname{Workshop on the CKM Unitarity Triangle, IPPP Durham, April
  2003}

\title{Supersymmetric contributions to the CP asymmetry of  
the $B \to \phi K_S$ and $B\to \eta' K_S$ }

\author{S. Khalil\addressmark{a,b} and E.Kou\addressmark{a}}

\address[a]{IPPP, Physics Department, Durham University, DH1 3LE,
Durham,~~U.~K.}
\address[b]{Ain Shams University, Faculty of Science, Cairo, 11566,
Egypt.}

\begin{document}
%%%%%%%%%%%%%%%%%%%%%%%%%%%%%%%%%%%%%%%%%%%%%%%%%%%%%
\begin{abstract}
We analyse the CP asymmetry of the $B \to \phi K_S$ and $B\to \eta' K_S$ processes 
in general supersymmetric models. We show that chromomagnetic type of operator may play an 
important role in accounting for the deviation of the mixing CP asymmetry between $B \to \phi 
K_S$ and $B \to J/\psi K_S$ processes observed by Belle and BABAR experiments.
We also show that due to the different parity in the final states of these processes, their 
supersymmetric contributions from the R-sector have an opposite sign, which naturally 
explain the large deviation between their asymmetries. 
\end{abstract}
\maketitle

%%%%%%%%%%%%%%%%%%%%%%%%%%%%%%%%%%%%%%%%%%%%%%%%%%%
%% standard LaTeX from here on...

\section{Introduction}
One of the most important tasks for B factory experiments would be to test  
the Kobayashi-Maskawa (KM) ansatz for the flavor CP violation. The flavor CP violation 
has been studied quite a while, however, it is still one of the least tested aspect
in the standard model (SM). Although it is unlikely that the SM provides the 
complete description of CP violation in nature (e.g. Baryon asymmetry in the universe), 
it is also very difficult to 
include any additional sources of CP violation beyond the phase in the CKM mixing 
matrix. Stringent constraints on these phases are usually obtained from the 
experimental bounds on the electric dipole moment (EDM) of the neutron, electron 
and mercury atom. Therefore, it remains a challenge for any new physics beyond the 
SM to give a new source of CP violation that may explain possible deviations 
from the SM results and also avoid overproduction of the EDMs. In supersymmetric 
theories, it has been emphasised ~\cite{Abel:2001vy} that there are attractive scenarios
where the EDM problem is solved and genuine SUSY CP violating effects are found. 

Recently, BABAR and Belle collaborations announced a $2.7 \sigma$ deviation from $\sin 2\beta$ in 
the $B \to \phi K_S$ process \cite{belle,babar}. 
In the SM, the decay process of $B \to \phi K$ is dominated by the top quark 
intermediated penguin 
diagram, which do not include any CP violating phase. Therefore,  
the CP asymmetry of $B \to J/\psi K_S$ and $B \to \phi K_S$ in SM are caused 
only by the phase in $B^0-\overline{B}^0$ mixing diagram and we expect 
$S_{J/\psi K_S} = S_{\phi K_S}$ where $S_{f_{CP}}$ represents the mixing CP asymmetry. 
The $B \to \ep K_S$ process is induced by more diagrams since $\ep$ meson contains not only 
$s\bar{s}$ state but also $u\bar{u}$ and $d\bar{d}$ states with the pseudoscalar mixing 
angle $\theta_p$. Nevertheless, under an assumption that its tree diagram contribution 
is very small, which is indeed the case, one can expect $S_{\phi K_S} =S_{\ep K_S}$ 
\cite{belle,babar2} as well. 
Thus, the series of new experimental data surprised us: 
\begin{eqnarray}
&&S^{\mbox{\tiny exp.}}_{\tiny {J/\psi K_S}} = 0.734\pm 0.054, \label{eq:1} \\
&&S^{\mbox{\tiny exp.}}_{\tiny {\phi K_S}} = -0.39\pm 0.41, \label{eq:2} \\
&&S^{\mbox{\tiny exp.}}_{\tiny {\eta^{\prime} K_S}} = 0.33\pm 0.41 \label{eq:3} 
\end{eqnarray}

It was pointed out \cite{Nir} that the discrepancy between Eq. (\ref{eq:1}) and Eq. (\ref{eq:2})
might be explained by new physics contribution through the penguin diagram to $B \to \phi K_S$. 
However, in that case, a simultaneous explanation for the discrepancy 
between Eq. (\ref{eq:2}) and Eq. (\ref{eq:3}) is also necessary. We show our attempts to 
understand all the above experimental data within the Supersymmetric models.  

%%%%%%%%%%%%%%%%%%%%%%%%%%%%%%%%%%%%%%%%%%%%%%%%%%%%%%%%%%%%
\section{The mass insertion approximation}
As mentioned, the SUSY extension of the SM may provide considerable effects to 
the CP violation observables since it contains new CP violating phases and also new
flavour structures. Thus, SUSY is a natural candidate to resolve the discrepancy among the 
observed mixing CP asymmetries in $B$-meson decays.
 
In the following, we will perform a model independent 
analysis by using the mass insertion approximation \cite{HallRaby}. We
start with the minimal supersymmetric standard model (MSSM),
where a minimal number of super-fields is introduced and $R$ parity
is conserved, with the following soft SUSY breaking terms
\bea\label{susy:gen:vsb}
V_{SB} &=& m_{0\alpha}^2 \phi_{\alpha}^* \phi_{\alpha} +
\epsilon_{ab}
\Big(A^u_{ij} Y^u_{ij} H_2^b \tilde{q}_{L_i}^a \tilde{u}^*_{R_j} +
A^d_{ij} Y^d_{ij} H_1^a \tilde{q}_{L_i}^b \tilde{d}^*_{R_j} \nonumber\\
&+& A^l_{ij} Y^l_{ij} H_1^a \tilde{l}_{L_i}^b \tilde{e}^*_{R_j} - B\mu H_1^a H_2^b + 
\mathrm{H.c.} \Big)\nonumber\\
&-& \frac{1}{2} \Big(m_3\bar{\tilde{g}} \tilde{g} +
m_2 \overline{\widetilde{W^a}} \widetilde{W}^a +
m_1 \bar{\tilde{B}} \tilde{B}\Big)\;,
\eea
where $i,j$ are family indices, $a,b$ are $SU(2)$ indices, and
$\epsilon_{ab}$ is the $2\times 2$ fully antisymmetric tensor, with
$\epsilon_{12}=1$. Moreover, $\phi_{\alpha}$ denotes all
the scalar fields of the theory.
Although in general the parameters $\mu$,
$B$, $A^\alpha$ and $m_i$ can be complex, two of their
phases can be rotated away.

The mass insertion approximation is a technique which is developed to include the 
soft SUSY breaking term without specifying the models in behind. In this approximation, 
one adopts a basis where the couplings of the fermion and sfermion are flavour diagonal, leaving
all the sources of flavour violation inside the off-diagonal terms of the sfermion mass
matrix. These terms are denoted by $(\Delta^{q}_{AB})^{ij}$,
where $A,B=(L,R)$ and $q=u,d$. The sfermion propagator
is then expanded as 
\begin{equation}
\langle \tilde q_A^a \tilde q_B^{b*} \rangle ={\rm i}~(k^2 {\bf 1}-\tilde{m}^2 {\bf 1}- 
\Delta^q_{AB})^{-1}_{a b}\simeq
{{\rm i}~\delta_{ab}\over k^2-\tilde{m}^2}  +{{\rm i}~(\Delta^q_{AB})_{ab}\over 
(k^2-\tilde{m}^2)^2},
\end{equation}
where ${\bf 1}$ is the unit matrix and $\tilde{m}$ is the average squark mass.
The SUSY contributions
are parameterised in terms of the dimensionless parameters
 $(\delta^{q}_{A B})_{ij}=
(\Delta^{q}_{AB})^{ij}/\tilde{m}^2$.
This method allows to parametrise, in a
model independent way, the main sources of flavor violations in SUSY models.

Including the SUSY contribution, the effective Hamiltonian for the penguin diagrams 
are written as
\begin{equation}
\mathcal{H}^{\Delta B=1}_{\mbox{eff}}\!=\!-\frac{G_F}{\sqrt{2}}V_{tb}V_{ts}^*
\left[\sum_{i=3}^{6}C_iO_i\!+\!C_gO_g \!+\!
\sum_{i=3}^{6}\tilde{C}_i\tilde{O}_i\!+\!\tilde{C}_g\tilde{O}_g\right] 
\label{eq:4}
\end{equation}
where $C_3\sim C_6 (\tilde{C}_3\sim \tilde{C}_6)$ include  $(\delta_{LL})_{23} ((\delta_{RR})_{23})$ contributions and 
$C_g (\tilde{C}_g)$ include $(\delta_{LL, LR})_{23} ((\delta_{RR, RL})_{23}$). 
The terms with tilde are obtained from $C_{i,g}$ and $O_{i,g}$ by exchanging 
$L \leftrightarrow R$.

\begin{center}
\begin{figure}[h]
\includegraphics[width=8cm]{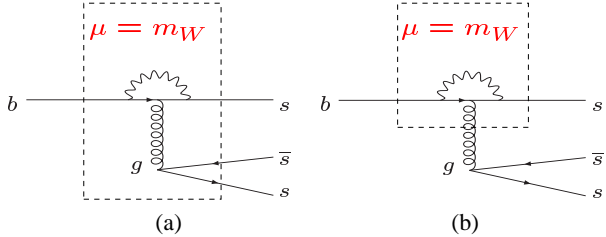}\\
\small{\hspace*{2cm}(a) \hspace{3.5cm}(b)}
\caption{(a)$O_3- O_6$ contributions which include $(\delta_{LL, RR})_{23}$ mass insertions. 
(b) $O_g$ contribution which includes 
$(\delta_{LL, RR, LR, RL})_{23}$ mass insertions. }
\label{fig:1}
\end{figure}
\end{center}

As emphasised in \cite{KK}, the leading contribution to $\Delta B=1$ processes come from 
the chromomagnetic penguin operator $O_{g} (\tilde{O}_g)$, in particular from the part proportional to 
the LR (RL) mass insertions which is enhanced by a factor $m_{\tilde{g}}/m_b$, where $C_g$ 
$\tilde{C}_g$ are given by
\be
C_g \sim \frac{\alpha_s \pi}{\tilde{m}} \frac{m_{\tilde{g}}}{m_b} (\delta^d_{23})_{LR} 
\ \ \ \ \ 
\tilde{C}_g \sim \frac{\alpha_s \pi}{\tilde{m}} \frac{m_{\tilde{g}}}{m_b} (\delta^d_{23})_{RL}.
\ee
Note that the mass insertions appearing in the box diagrams are $(\delta_{AB})_{13}$ 
($A,B = L\  \mbox{or}\  R$), thus,  SUSY contributions to box diagram and to penguin diagram 
are independent. $S_{J/\psi}\simeq \sin 2\beta$ indicates the smallness of 
$(\delta_{AB})_{13}$ \cite{GK}.
%%%%%%%%%%%%%%%%%%%%%%%%%%%%%%%%%%%%%%%%%%%%%%%%%%%%%%%%%%%
\section{Can we explain the experimental data of $S_{\phi K_S}$ in SUSY? }
Following the parametrisation of the SM and SUSY amplitudes in Ref.\cite{KK},
$S_{\phi K_S}$ can be written as
\begin{eqnarray}
S_{\phi K_S}=\Frac{\sin 2 \beta +2 R_{\phi}
\cos \delta_{12} \sin(\theta_{\phi} + 2 \beta) +
R_{\phi}^2 \sin (2 \theta_{\phi} + 2 \beta)}{1+ 2 R_{\phi}
\cos \delta_{12} \cos\theta_{\phi} +R_{\phi}^2}
\end{eqnarray}
where $ R_{\phi}= \vert A^{\susy}/A^{\sm}\vert$, $\theta_{\phi}=
\mathrm{arg}(A^{\susy}/A^{\sm})$, and $\delta_{12}$ is the strong phase.

We will discuss in the following whether the SUSY contributions can 
make $S_{\phi K_S}$ negative.  
Note that the mass insertions $(\delta_{AB})_{23}$ have already been 
constrained by the experimental data for $Br(B \to X_s \gamma)$ :  
\begin{equation}|(\delta_{LL, RR})_{23}| \leq 1, \ \ \ \ \ \ 
|(\delta_{LR, RL})_{23}| \leq 1.6\times 10^{-2} \label{eq:6}\end{equation}

For $m_{\tilde{q}}=m_{\tilde{g}}=500$\ GeV, we obtain  
\small
\begin{equation}
\frac{A^{\mathrm{SUSY}}}{A^{\mathrm{SM}}}
=0.23(\delta_{LL})_{23}+0.23(\delta_{RR})_{23}+97.4(\delta_{LR})_{23}+97.4(\delta_{RL})_{23}
\label{eq:7}
\end{equation}
\normalsize
The constrains from $Br(B \to X_s \gamma)$ gives the maximum $|A^{\mathrm{SUSY}}|/|A^{\mathrm{SM}}|$: 
\begin{eqnarray}
\frac{|A^{\mathrm{SUSY}}|}{|A^{\mathrm{SM}}|}&<& 0.23 \ \ \ \ \ 
\mbox{For}\ \ (\delta_{LL, RR})_{23} \label{eq:8}\\
\frac{|A^{\mathrm{SUSY}}|}{|A^{\mathrm{SM}}|}&<& 1.6 \ \ \ \ \ 
\mbox{For}\ \ (\delta_{LR, RL})_{23} \label{eq:9}
\end{eqnarray}
In Fig.{\ref{fig:2}}, we present plots for 
the phase of $(\delta_{LL(RR)}^d)_{23}$ and $(\delta_{LR(RL)}^d)_{23}$ versus
the mixing CP asymmetry $S_{\phi K_S}$ when the strong phases are ignored. 
We choose the three values of the magnitude of these mass 
insertions within the 
bounds from the experimental limits from  $B \to X_s \gamma$.  
Each plot shows a contribution from an individual mass insertion by setting the 
other three to be zero. 
As can be seen from these plots, the $LR$ (same for $RL$) gives the largest contribution 
to $S_{\phi K_S}$.  In order to have 
a sizable effect from the $LL$ or $RR$, 
the magnitude of  $(\delta_{LL(RR)}^d)_{23}$ has to be of order one 
and furthermore, the imaginary part needs to be as large as the real part.   
In any case, it is very difficult to give negative value of $S_{\phi K_S}$ from $(\delta_{LL}^d)_{23}$ or $(\delta_{RR}^d)_{23}$ mass insertion.  
If the experimental data remains as small as the current values, 
$LL, RR$  dominated models would get  sever constrains on some parameters. 
Note that $S_{\phi K_S}$ decreases as SUSY masses becomes smaller. 
In \cite{CFMS}, a choice of 
$m_{\tilde{q}}\simeq 350$ GeV has been used and a negative $S_{\phi K_S}$ for $LL, RR$ models has been obtained. 
\begin{center}\hspace*{-1cm}
\begin{figure}[h]
\psfrag{(a)}[l][l][0.8]{(a)\ \ $|(\delta_{LL(RR)}^d)_{23}|$}
\psfrag{(c)}[l][l][0.8]{(b)\ \ $|(\delta_{LR(RL)}^d)_{23}|$}
\psfrag{s}[r][l][0.8]{$S_{\phi K_S}$}
\psfrag{1ll}[l][l][0.5]{\Large $0.1$}
\psfrag{2ll}[l][l][0.5]{\Large $0.5$}
\psfrag{3ll}[l][l][0.5]{\Large $\ \ 1$}
\psfrag{1lr}[l][l][0.5]{\Large $0.001$}
\psfrag{2lr}[l][l][0.5]{\Large $0.005$}
\psfrag{3lr}[l][l][0.5]{\Large $0.01$}
\psfrag{ll}[l][l][0.6]{$\arg[(\delta_{LL(RR)}^d)_{23}]$}
\psfrag{lr}[l][l][0.6]{$\arg[(\delta_{LR(RL)}^d)_{23}]$}
\includegraphics[width=8cm]{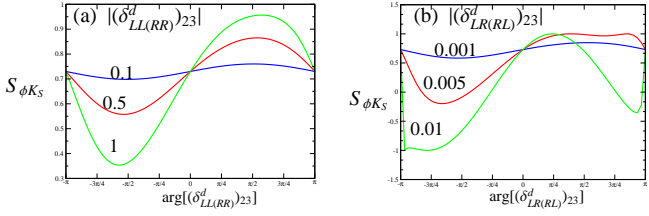}\\
\caption{Result for $S_{\phi K_S}$ in terms of the phase in mass insertions.}
\label{fig:2}
\end{figure}
\end{center}
%%%%%%%%%%%%%%%%%%%%%%%%%%%%%%%%%%%%%%%%%%%%%%%%%%%%%%
\section{What happened to the $B \to \ep K_S$ process? }
Although $B \to \phi K_S$ and $B \to \eta^{\prime} K_S$ are very similar processes, 
the parity of the final states can deviate the result. 
In the naive factorisation approximation, the amplitudes are written as a product 
of Wilson coefficients, form factors and decay constants:  
\begin{equation}
A(B \to \phi (\ep) K)\propto C_{\mbox{\tiny Wilson}}\ F^{B\to K} f_{\phi(\ep)} 
\end{equation}
The decay constants appear in the calculation by sandwiching the $V\pm A$ current 
($O_i$ and $\tilde{O}_i$ contributions, respectively) with $\phi(\ep)$ and vacuum: 
\begin{eqnarray}
\langle 0|\overline{s} \gamma_{\mu} (1\pm \gamma_5)s|\phi\rangle &=&
m_{\phi}f_{\phi}\epsilon_{\mu} \label{eq:11} \\
\langle 0|\overline{s}  \gamma_{\mu} (1\pm \gamma_5)s|\eta^{\prime}\rangle &=&
\pm if_{\eta^{\prime}}p_{\mu} \label{eq:12}
\end{eqnarray}
As can be seen from Eqs. (\ref{eq:11}) and (\ref{eq:12}), the vector meson $\phi$ 
picks up the $\gamma_{\mu}$ term while the pseudoscalar meson $\ep$ picks up the 
$\gamma_{\mu}\gamma_5$ term so that contributions from $O_i$ and $\tilde{O}_i$ obtain  
opposite signs for $\ep$. 
 
\begin{center}
\begin{figure}[h]
\includegraphics[width=4cm]{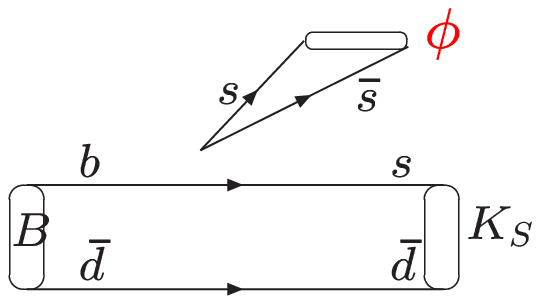}
\includegraphics[width=4cm]{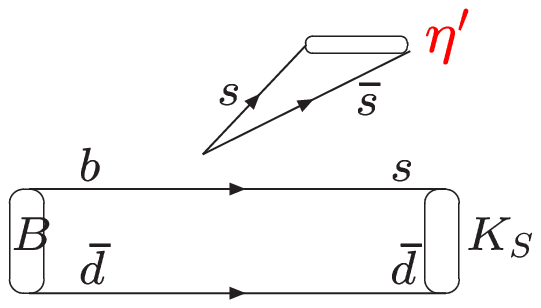}\\
\caption{Schematically described naive factorisation approximation for 
$B \to \phi K_S$ and $B \to \ep K_S$ processes.}
\label{fig:3}
\end{figure}
\end{center}

\begin{center}
\begin{figure}[h]
\psfrag{phi}[l][l]{$S_{\phi K_S}^{\mbox{\tiny exp.}}$}
\psfrag{eta}[l][l]{$S_{\eta^{\prime} K_S}^{\mbox{\tiny exp.}}$}
\psfrag{s}[l][l]{$S_{\phi K_S, \eta^{\prime} K_S}$}
\psfrag{t}[l][l]{$\theta_{\phi K_S}$}
\includegraphics[width=8cm]{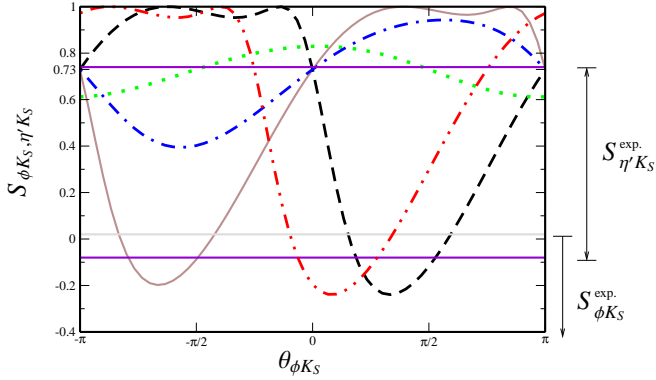}
\caption{Case 1: $|\delta_R|$ dominating.  
Case 2: $|\delta_R|=|\delta_L|$ with $\Delta\theta=\pi/10$. 
Case 3: $\arg\delta_L=\arg\delta_R$ with $\Delta|\delta|=+0.2$.  
Case 4: $\arg\delta_R=\arg\delta_L+\pi/2$ with $\alpha=3\pi/8$. 
Solid line: $S_{\phi K_S}$ for  $|A^{\mathrm{SUSY}}|/|A^{\mathrm{SM}}|=0.5$. }
\label{fig:4}
\end{figure}
\end{center}

As a result, the sign of the $RR$ and $RL$ contributions are different for 
$B \to \phi K_S$ and $B \to \ep K_S$ \cite{KK2}: 
\small
\begin{eqnarray*}
\!\!\left(\frac{A^{\mathrm{SUSY}}}{A^{\mathrm{SM}}}\right)_{\phi K_S}
\!\!\!\!\!&\!=\!\!\!&\!0.23(\delta_{LL})_{23}\!+\!0.23(\delta_{RR})_{23}
\!+\!97.4(\delta_{LR})_{23}\!+\!97.4(\delta_{RL})_{23}\label{eq:13} \\
\!\!\left(\frac{A^{\mathrm{SUSY}}}{A^{\mathrm{SM}}} \right)_{\eta^{\prime} K_S}
\!\!\!\!&\!=\!\!\!&\!0.23(\delta_{LL})_{23}\!-\!0.23(\delta_{RR})_{23}
\!+\!101(\delta_{LR})_{23}\!-\!101(\delta_{RL})_{23}\label{eq:14}
\end{eqnarray*}
\normalsize
Since the coefficient for each mass insertions are similar, 
 we use the following definition to simplify our following discussions: 
\begin{equation}
\left(\frac{A^{\mathrm{SUSY}}}{A^{\mathrm{SM}}} \right)_{\phi K_S}
\equiv \delta_L+\delta_R \ \ \ \ 
\left(\frac{A^{\mathrm{SUSY}}}{A^{\mathrm{SM}}}\right)_{\eta^{\prime} K_S}
\equiv  \delta_L-\delta_R\label{eq:15}\end{equation}
where $\delta_L$ includes contributions from $(\delta_{LL})_{23}$ and $(\delta_{LR})_{23}$ and 
$\delta_R$ includes contributions from $(\delta_{RR})_{23}$ and $(\delta_{RL})_{23}$. 
Now let us show how this sign flip effects to the mixing CP violation $S_{\phi KS}$ 
and $S_{\ep K_S}$. Since both $\delta_L$ and $\delta_R$ are complex number, we have 
four parameters to be fixed while we have only two experimental data. Thus, we fix two 
parameters and perform a case-by-case study in the following. 
\begin{itemize}
\item Case 1: $|\delta_R|\gg|\delta_L|$ 
\begin{eqnarray}
\left(\frac{A^{\mathrm{SUSY}}}{A^{\mathrm{SM}}} \right)_{\phi K_S}&=&|\delta_R|e^{i\arg\delta_R}, \nonumber \\
\left(\frac{A^{\mathrm{SUSY}}}{A^{\mathrm{SM}}}\right)_{\eta^{\prime} K_S}&=&|\delta_R|e^{i(\arg\delta_R+\pi)}.\nonumber 
\end{eqnarray}
\item Case 2: $|\delta_L|=|\delta_R|$ ($\Delta\theta=\arg\delta_L-\arg\delta_R$)
\begin{eqnarray}
\left(\frac{A^{\mathrm{SUSY}}}{A^{\mathrm{SM}}} \right)_{\phi K_S}&=& 2|\delta_L|\cos\frac{\Delta\theta}{2}
e^{i(\arg\delta_L+\arg\delta_R)/2}  \nonumber \\
\left(\frac{A^{\mathrm{SUSY}}}{A^{\mathrm{SM}}}\right)_{\eta^{\prime} K_S}&=& 2|\delta_L|\sin\frac{\Delta\theta}{2}
e^{i(\arg\delta_L+\arg\delta_R+\pi)/2}\nonumber 
\end{eqnarray}
\item Case 3: $\arg\delta_L=\arg\delta_R$ ($\Delta|\delta|=|\delta_L|-|\delta_R|$)\\
\begin{eqnarray}
\left(\frac{A^{\mathrm{SUSY}}}{A^{\mathrm{SM}}} \right)_{\phi K_S}&=&(|\delta_L|+|\delta_R|)e^{i\arg\delta_L} \nonumber \\
\left(\frac{A^{\mathrm{SUSY}}}{A^{\mathrm{SM}}}\right)_{\eta^{\prime} K_S}&=&\Delta|\delta|e^{i\arg\delta_L}\nonumber 
\end{eqnarray}
\item Case 4:  $\arg\delta_R=\arg\delta_L+\pi/2$ ($\tan\alpha =|\delta_R|/|\delta_L|$)\\
\begin{eqnarray}
\left(\frac{A^{\mathrm{SUSY}}}{A^{\mathrm{SM}}} \right)_{\phi K_S}&=&\sqrt{|\delta_L|^2+|\delta_R|^2}
e^{i(\arg\delta_L+\alpha)} \nonumber \\
\left(\frac{A^{\mathrm{SUSY}}}{A^{\mathrm{SM}}}\right)_{\eta^{\prime} K_S}&=&\sqrt{|\delta_L|^2+|\delta_R|^2}
e^{i(\arg\delta_L-\alpha)}\nonumber 
\end{eqnarray}
\end{itemize}
In Fig. \ref{fig:4}, we show some examples of the parameter sets with which 
we can reproduce both experimental data of $S_{\phi K_S}$ and $S_{\ep K_S}$. 

%%%%%%%%%%%%%%%%%%%%%%%%%%%%%%%%%%%%%%%%%%%%%%%%%%%
\section{On the branching ratio of $B \to \ep K_S$: Gluonium vs. New physics}
In 1997, CLEO collaboration reported an unexpectedly large branching ratio \cite{CLEO}
\begin{equation}
Br^{\mbox{\tiny exp.}}(B^0\to K^0 \eta^{\prime}) =  
(89^{+18}_{-16}\pm 9)\times 10^{-6} 
\end{equation}
which is confirmed by Belle \cite{belle2} and BABAR \cite{babar2}: 
\begin{eqnarray}
\mbox{BELLE}&=&(79^{+12}_{-16}\pm 8)\times 10^{-6}, \\
\mbox{BABAR}&=&(76.9 \pm 3.5\pm 4.4)\times 10^{-6} 
\end{eqnarray}
Considering the theoretical prediction by the naive factorisation approximation
\begin{equation}
Br^{\mbox{\tiny theo.}}(B\to K \eta^{\prime}) \simeq 25\times 10^{-6}, 
\end{equation}
the experimental data is about factor of three large, thus, there have been 
various efforts to explain this puzzle.  
On one hand, new physics contributions have been discussed \cite{Kundu}. However, the enhancement 
by new physics contributions through penguin diagrams ends up with large branching 
ratios for all other penguin dominated processes. Therefore,  one needs a careful treatment 
to enhance only $B \to \ep K$ process without changing the predictions for the other 
processes.  
On the other hand, since this kind of large branching ratio is observed only in 
$B \to \ep K$ process, the gluonium contributors which only exist in this process 
have been a very interesting candidate to solve the puzzle \cite{soni} \cite{emi} 
though the amount of 
gluonium in $\ep$ is not precisely known \cite{emi2}.  In this section, let us discuss 
the effect of our including SUSY contributions to the branching ratios for 
$B \to \phi K$ and $B \to \ep K$. 

Inclusion of the SUSY contributions modify the branching ratio as:
\[Br^{\mbox{\tiny SM + SUSY}}=Br^{\mbox{\tiny SM}}\times [1+2\cos\theta_{\mbox{\tiny SUSY}}R+R^2]
\]
where $R=|A^{\mathrm{SUSY}}|/|A^{\mathrm{SM}}|$. 
As we have shown, to achieve a negative value of $S_{\phi K_S}$, we need 
$\theta_{\mbox{\tiny SUSY}}\simeq -\pi/2$, which  
suppresses the leading SUSY contribution. As a result, for instance, $R=0.5$ leads to: 
\begin{equation}
Br(B \to \phi K_S)=(7.8\times 10^{-6}) \times 1.25 = 9.7\times 10^{-6}
\end{equation}
which is within the experimental data $(9.1\pm 2.6)\times 10^{-6}$. 
On the other hand, the phase for $B \to \eta^{\prime} K $ is different from 
the one for $\phi K_S$, as is discussed in the previous section. 
For instance, Case 2 gives us the maximum value of: 
\begin{equation}
Br(B \to \eta^{\prime} K_S)=(25\times 10^{-6}) \times 3.25 = 81\times 10^{-6}
\end{equation}
However, this kind of enhancement would appear all the other two pseudo-scalars channels (such as 
$B \to K \pi$) and might cause some problems. 

\begin{center}
\begin{figure}[h]
\includegraphics[width=8cm]{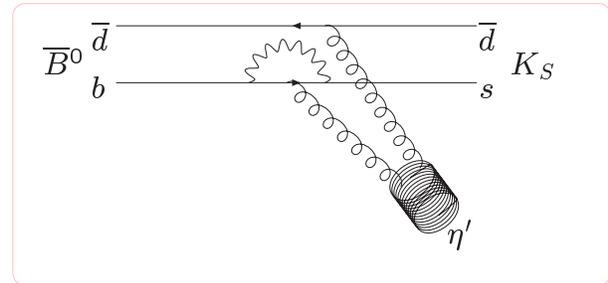}\\
\caption{A contribution from gluonium content in $\ep$ to the $B \to \ep K$ process. }
\label{fig:5}
\end{figure}
\end{center}

\begin{center}
\begin{figure}[h]
\small
\psfrag{S}[l][l][1.2]{$ S_{\ep K_S}$}
\psfrag{B}[l][l][1.1]{$ Br(B \to\ep K)/Br^{\sm}(B \to\ep K)$}
\psfrag{r}[l][l]{$ r=0 \rightarrow 0.3$}
\psfrag{alpha}[l][l][0.7]{$\arctan|\delta_R|/|\delta_L| =\pi/2 \rightarrow \pi/4$}
\psfrag{dt}[l][r][0.7]{$(arg\delta_L-arg\delta_R)=\pi/2 \rightarrow \pi/10$}
%\psfrag{dd}[l][l][0.7]{$|\delta_L|-|\delta_R|=-0.5 $}
%\psfrag{dd2}[l][l][0.7]{$\rightarrow 0.2$}
\psfrag{dd3}[l][l][0.7]{$|\delta_L|-|\delta_R|=-0.5 \rightarrow +0.2$}
\psfrag{c1}[l][l][1]{Case 1}
\psfrag{c2}[l][l][1]{Case 2}
\psfrag{c3}[l][l][1]{Case 3}
\psfrag{c4}[l][l][1]{Case 4}
\psfrag{sexp}[l][l][1.2]{$S_{\ep K_S}^{\mbox{\tiny exp}}$}
\includegraphics[width=8cm]{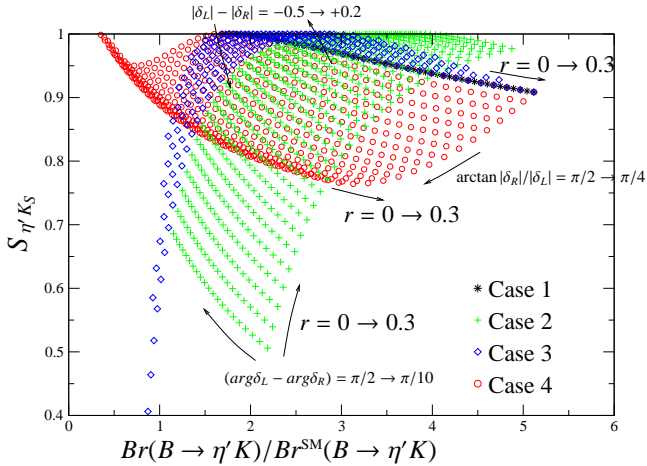}
\caption{Branching ratio versus the mixing CP asymmetry in the $B \to \ep K$ process. 
The parameter $r$ represents the contribution from the gluonium diagram.  }
\label{fig:6}
\end{figure}
\end{center}

As a whole, we would like to suggest that the solution for the branching ratio puzzle is 
not only the SUSY contribution but a combination of SUSY contribution and the 
gluonium contribution. Here, let us show the dependence of the gluonium contribution 
to the $S_{\ep K_S}$. Including the gluonium contribution (see Fig. \ref{fig:5}), 
the amplitude is modified to
\begin{equation}
A=A^{\sm}_{\ep K_S}+A^{\susy}_{\ep K_S}+G^{\sm}+G^{\susy} 
\end{equation}
where $G^{\sm}$ and $G^{\susy}$ are the new mechanism contributions to SM and 
SUSY, respectively.
Let us parametrise the unknown gluonium content in $\eta^{\prime}$ as 
$r=G^{\sm}/A^{\sm}_{\ep K_s}$. Our result is shown in Fig. \ref{fig:6} when we vary 
$r$ from 0 to 0.3.  As can be seen from this figure, the dependence of $S_{\ep K_s}$ on 
$r$ is not very strong, therefore, we can enhance the branching ratio by gluonium 
contribution without disturbing our findings for $S_{\ep K_S}$ in the previous section. 

%%%%%%%%%%%%%%%%%%%%%%%%%%%%%%%%%%%%%%%%%%%%%%%%%%%
\section{Conclusions}
We studied the supersymmetric contributions to the CP asymmetry of 
$B \to \phi K_S$ and $B \to \ep K_S$ in a model independent way. 
We found that the observed large discrepancy between $S_{J/\psi K_S}$ and 
$S_{\phi K_S}$ can be explained within some SUSY models with large $(\delta_{LR})_{23}$ 
or $(\delta_{RL})_{23}$ mass insertions. 
We showed that the SUSY contributions of $(\delta_{RR})_{23}$ and 
$(\delta_{RL})_{23}$ to  $B \to \phi K_S$ and $B \to \eta^{\prime} K_S$ 
have different signs. Therefore, the current observation,  
$S_{\phi K_S} < S_{\ep K_S}$, favours the $(\delta_{RR,RL})_{23}$ dominated models.  
We also discussed the SUSY contributions to the branching ratios. 
We showed that negative $S_{\phi K_S}$ and small SUSY effect to $Br(B \to \phi K)$ 
can be simultaneously achieved. On the other hand, we showed that 
SUSY contribution itself does {\it not} solve the puzzle of the large branching ratio 
of $B \to \eta^{\prime} K_S$. 
We included the gluonium contributions to $B \to \eta^{\prime} K_S$. We found 
that  our conclusion for $S_{\ep K_S}$does not disturbed by gluonium contributions. 
As soon as the experimental errors are reduced, the CP violation of 
$B \to \phi K_S$ and $B \to \ep K_S$ will be able to give a strong constraints 
on the $(\delta_{AB})_{23}$ mass insertions. 

%%%%%%%%%%%%%%%%%%%%%%%%%%%%%%%%%%%%%%%%%%%%%%%%%%%

\end{document}